\documentstyle[pre,aps,epsfig]{revtex}

\input{psfig}

\begin{document} 



\title{Using the Continuous Spectrum to ``Feel'' Integrability: the Effect
of Boundary Conditions}
\author{Panayotis G.\ Kevrekidis$^{\dagger}$ and 
Niurka R.\ Quintero$^\ddagger$} 
\address{$^{\dagger}$Department of Mathematics and Statistics, University of
Massachusetts, Amherst MA 01003-4515, USA; \\
$^{\ddagger}$Departamento de F\'{\i}sica Aplicada I,
Escuela Universitaria Polit\'ecnica, 
Universidad de Sevilla, Virgen de \'Africa 7, 41011, Sevilla,
Spain; and Instituto {\sc Carlos I} de F\'{\i}sica Te\'orica y
Computacional Universidad de Granada. E-18071 Granada, Spain}
\maketitle

\begin{abstract} 
The scope of this work is to propose a method for testing the
integrability of a model partial differential (PDE) and/or differential
difference equation (DDE). For monoparametric families of PDE/DDE's,
that are known to possess isolated integrable points, we find that
very special features occur in the continuous (``phonon'') 
spectrum at these ``singular''
points. We identify these features in the case example
of a PDE and a DDE (that sustain front and pulse-like 
solutions respectively) for different types of boundary
conditions. The key finding of the work is that such spectral features
are generic near the singular, integrable points and hence we 
propose to explore a given PDE/DDE for such traits, as a means
of assessing its potential integrability.
\end{abstract}

\vspace{2mm}

Integrable models of partial differential (PDE) and differential difference
(DDE) equations have been a topic of intense investigation over the past
few decades \cite{AS,Gibbon,scott}. The main reason for that, except for the
wide variety of physical applications that can be described by integrable
or near-integrable systems, is that the special case of integrable models
can be analyzed completely by means of the inverse scattering transform
\cite{AS,FAD}. This can then serve as a starting point for perturbative
treatment of near-integrable systems.

In the process of these developments, a number of techniques have been
developed for assessing integrability in continuous \cite{h3} or
discrete \cite{h4} settings (or applicable to both \cite{h5}). An interesting
feature of these ``tests'' 
is that they are necessary (but not sufficient)
conditions for integrability. Hence, if a model equation fails such a 
criterion, it is non-integrable, but if it passes, it may or {\it may not}
be integrable. In a sense, this suggests that we still do not understand
the {\it essential ingredients} that render a system completely 
integrable. Of course, should a Lax pair be identified and the inverse
scattering mechanism be applied, we know that the system is integrable,
but it would certainly be desirable (as is clear from all the above 
effort to create ``integrability tests'') to have a mechanistic 
(``black box'') type of criterion to assess that.

We, of course, do not claim to be providing a full answer to this
question in the present work. However, we will try to give a 
number of useful hints that may lead to partial answers to the above
questions and may provide some new intuition in the effort to construct
such mechanistic criteria.

Our tool of choice will be the use of different sets of boundary
conditions (BC) to ``feel'' the continuous spectrum (CS) of the linearization
around the nonlinear coherent structure that the PDE/DDE of
interest supports. Notice that the effect of boundary conditions
in related contexts has been studied in a number of references;
see e.g., \cite{tru} and references therein. However, in all
of these works the effects of the BC to the point spectrum
(rather than the CS) were assessed and moreover this was not
done in direct connection with issues of integrability.

In the present work, we focus on two model problems, to 
establish our findings and demonstrate their generality.
The models are selected as one parameter families of
equations such that one member of the family is an integrable
system. Moreover, in illustrating the generality of the 
conclusions, they are selected in a form such that the
one model corresponds to a PDE, while the other to a DDE
and so that the one is kink 
bearing, while the other is pulse 
bearing. The models of interest will be the parametrically modified
sine-Gordon equation (often also called the Peyrard-Remoissenet (PR) model)
\cite{prb} and a modified version of the discrete nonlinear
Schr{\"o}dinger (DNLS) model (occasionally
called the Salerno model) \cite{Cai}. 
The former PDE reads:
\begin{eqnarray}   \label{eq1}
\phi_{tt} - \phi_{xx} & = & - \frac{dU}{d \phi}, 
\quad  \quad 
U(\phi,r) = \frac{(1-r)^{2} \, [1-\cos(\phi)]}
{1+r^{2}+2 r \cos(\phi)}, \quad  |r| < 1, \quad 
|x| < \infty, 
\end{eqnarray} 
while the latter DDE is of the form:
\begin{eqnarray}
i \dot{u}_n=-\Delta_2 u_n - |u_n|^2 \left[ 2 \epsilon u_n + (1-\epsilon) 
(u_{n+1}+u_{n-1}) \right].
\label{neq2}
\end{eqnarray}
The most well-known among these monoparametric families of models
are the sine-Gordon equation (Eq. (\ref{eq1}), for $r=0$) which is
relevant to superconductivity and charge density waves among other
applications \cite{Gibbon} and the experimentally realizable 
discrete nonlinear Schr{\"o}dinger equation \cite{b7} of $\epsilon=1$,
as well as its integrable, so-called Ablowitz-Ladik \cite{AL} counterpart
for $\epsilon=0$ in the case of Eq. (\ref{neq2}).

Notice that for the PDE, the subscripts denote partial derivatives
of the field, while for the DDE, the overdot denotes temporal
derivative, $\Delta_2 u_n \equiv C (u_{n+1} - 2 u_{n} + u_{n-1})$ 
where $C = 1/(\Delta x)^{2}$ is a constant determined by the 
lattice spacing
$\Delta x$; 
the subscript $n$ denotes the lattice site index. 
In the former case, there
exist kink-like solutions which have been detailed in \cite{prb},
while in the latter, the field is complex and 
there exist pulse-like solutions of the form
$u_n= \exp(i \Lambda t) v_n$, where $\Lambda$ is the frequency of
the solutions and $v_n$ its (real) exponentially localized spatial
profile \cite{Cai,b7}.

In the PDE, linearization around a state $\phi_0(x)$, using the
ansatz $\phi_{0}(x) + \delta \exp(i \omega t) f(x)$ into Eq.\ (\ref{eq1}),
yields to $O(\delta)$ the linearization equation
\begin{eqnarray} \label{eq3} 
f_{xx} + \left(\omega^{2} - \frac{(r^{2}-1)^{2} [(r^{2} + 1 ) 
\cos[\phi_{0}(x)] + r (3 - \cos[2 \phi_{0}(x)])]}
{[1 + r^{2} + 2 r \cos(\phi_{0}(x))]^{3}} \right) f & = & 0, \quad 
|x| < \infty.  
\end{eqnarray} 
Notice that when $r=0$, $\phi_{0}(x)= 4\arctan[\exp(x)]$ 
(static kink solution of 
the sG equation) and for this function, 
the Sturm-Liouville problem (\ref{eq3}) can be 
exactly solved \cite{rubi} yielding one discrete mode 
(Goldstone mode) at $\omega=0$ and the 
continuous spectrum represented by the phonons.
For $r \neq 0$, neither the static solution nor 
the linearization spectrum are 
explicitly available. Hence, 
to investigate the linear spectrum 
of  Eq.\ (\ref{eq3}), we should find its numerical solution,
by discretizing the equation in a numerical mesh for a finite domain.
The mesh consists of the $N+1$ points
$x_{j} =\{-L/2 + j \Delta x; \quad j=0,1,2,...,N  \}$ defined in the 
finite length $L$ of the system ($\Delta x=L/N$). 
Notice that since, in this case, we wish to 
emulate the behavior of the PDE, $\Delta x$ is very fine (typically
$0.05$), and the robustness of the findings upon variation of the
(small) $\Delta x$ has been verified. 

When we compute the solution either of the PDE or of the linearization
equation,
we consider three different types of BC. 
Free BC are expressed through $ f(0)=f(1)$ and $f(N)=f(N-1)$
(notice that $f(j)=f(x_j)$). Fixed BC are given by 
$f(0)=0$ and $f(N)=0$, while (anti-)periodic BC are given
by $f(0)=-f(N-1)$ and  $f(N)=-f(1)$. 

Analogously to the PDE, for the linear stalibity analysis 
of DDE (\ref{neq2}) we insert 
$\exp(i \Lambda t) [v_n + \delta (U_{n} {\rm e}^{- i \omega t} + 
W_{n} {\rm e}^{i \omega^{\star} t})]$ into Eq.\ (\ref{neq2}). We
thus obtain to
 order $O(\delta)$  
the following eigenvalue problem for 
$\{\omega, \{U_{n}, W_{n}^{\star}\}\}$ 
\begin{eqnarray} \label{ls2}
\omega \left(
\begin{array}{l}
U_{n} \\
W_{n}^{\star}
\end{array}
\right) & = & {\cal L} \left(
\begin{array}{l}
U_{n} \\
W_{n}^{\star}
\end{array}
\right), \qquad 
{\cal L} = \left(
\begin{array}{c|c}
A & B \\
\hline
-B & -A 
\end{array}
\right), 
\\
A_{mn} & = & [\Lambda + 2 C - (4 \epsilon v_{n}^{2} +(1-\epsilon) v_{n} 
[v_{n+1} + v_{n-1}])] \delta_{m,n} + 
[(1-\epsilon) v_{n}^{2} - C] (\delta_{m,n+1} + 
\delta_{m,n-1}),
\nonumber
\\
B_{mn} & = & - v_{n} [2 \epsilon v_{n} + 
(1-\epsilon) (v_{n+1} + v_{n-1})] \delta_{m,n},
\nonumber
\end{eqnarray}
where the stars denote complex conjugation. We compute 
the solutions of DDE (\ref{neq2}) and Eq.\ (\ref{ls2}) using $200$ points,  
$\Delta x = 0.75$. The BC are defined analogously through
$ U_{0} =  U_{1}$, $U_{N}=U_{N-1}$,
$W_{0} = W_{1}$, and $ W_{N} = W_{N-1}$ for free BC.
For fixed BC: $U_{0}  =  0$, $U_{N}=0$,
$W_{0} = 0$, and $W_{N}=0$, while for periodic BC:
$U_{0} = U_{N-1}$, $U_{N}=U_{1}$, 
$W_{0} = W_{N-1}$, and  $W_{N}=W_{1}$.

We have also tried no-flux boundary conditions that yield essentially
the same results as in the free BC case.

\begin{figure}[h]
\psfig{file=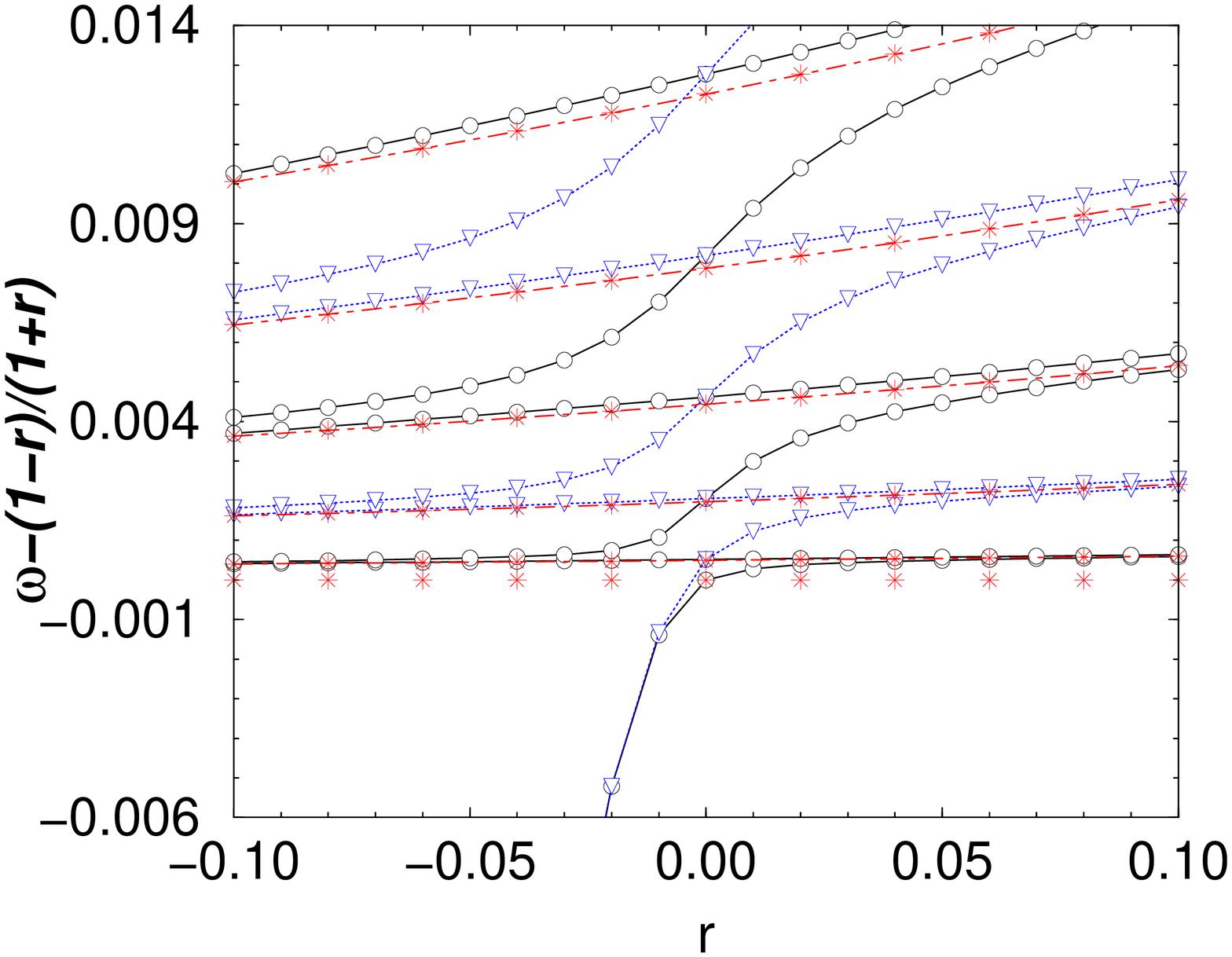,width=1.95in,angle=-90}
\psfig{file=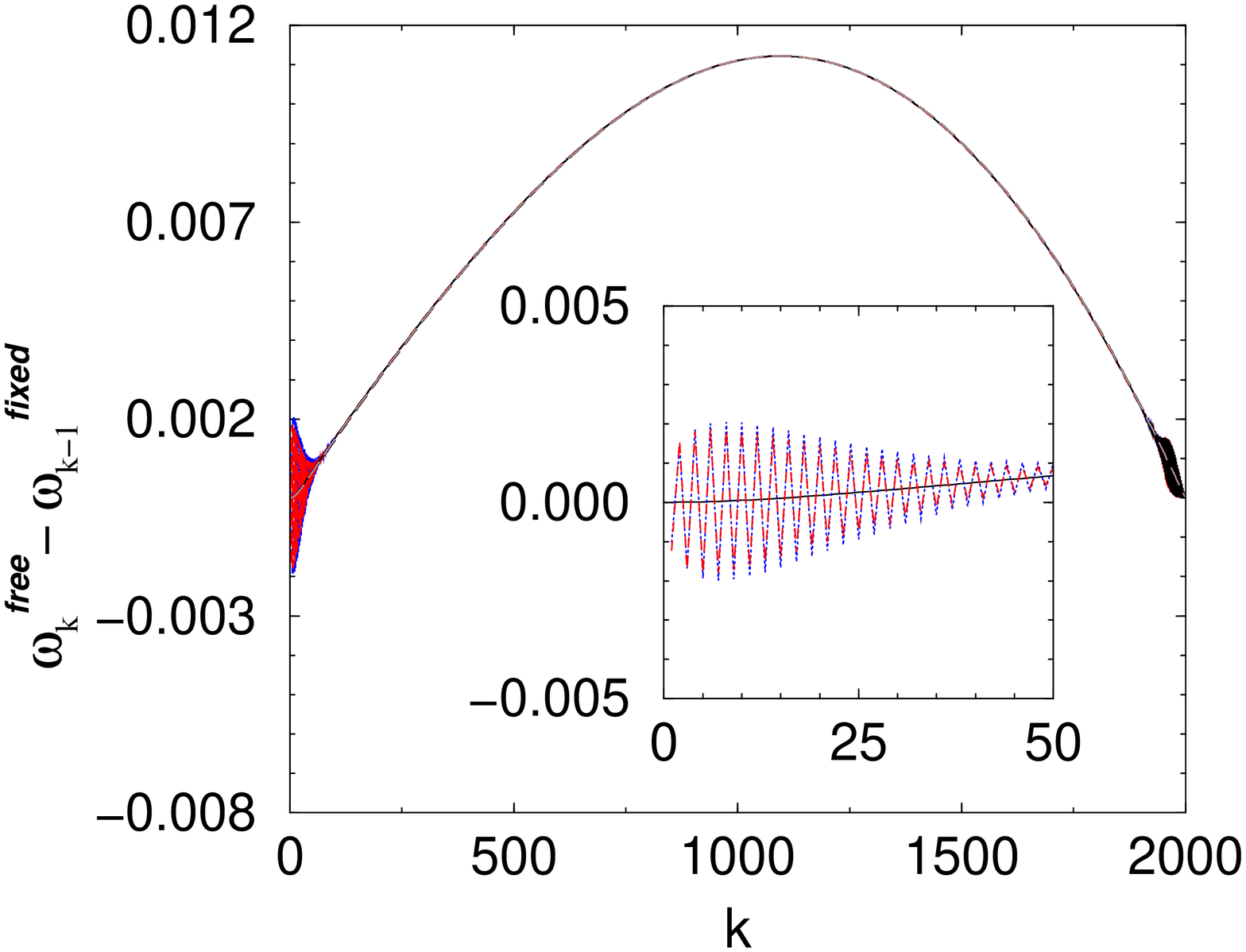,width=1.95in,angle=-90}
\caption{Comparison (left subplot) of the eigenfrequencies of 
linear spectrum for fixed and free BC: 
We have plotted the difference between the eigenfrequencies 
computed from the Eq.\ (\ref{eq3})  
and $\omega_{min}=(1-r)/(1+r)$ versus 
$r$. The circles joined by solid line (free BC)  
represent how far the frequencies are from the lower phonon mode. 
The triangles joined by dotted lines correspond to fixed BC.
Here, we have also added the distribution of the eigenfrequencies 
when we linearize around the zero solution by using stars (free BC) 
and dot-dashed line (fixed). The difference between the 
frequencies for free and fixed BC,  
$\omega_{k}^{free}-\omega_{k-1}^{fixed}$ ($2 \le k \le N-1$),  
is plotted in the right subplot as a function of the wave number 
for $r=0$ (solid line), $r=-0.02$ (dashed line) and 
$r=0.02$ (dotted line). 
We observe an oscillatory 
behaviour for the last wave numbers (in the inset the same 
difference is shown for the first wave numbers). 
For the linear PR case, the results essentially coincide with the 
integrable nonlinear case of $r=0$
(solid line).
}
\label{nfig1}
\end{figure}

\begin{figure}[!p]
\psfig{file=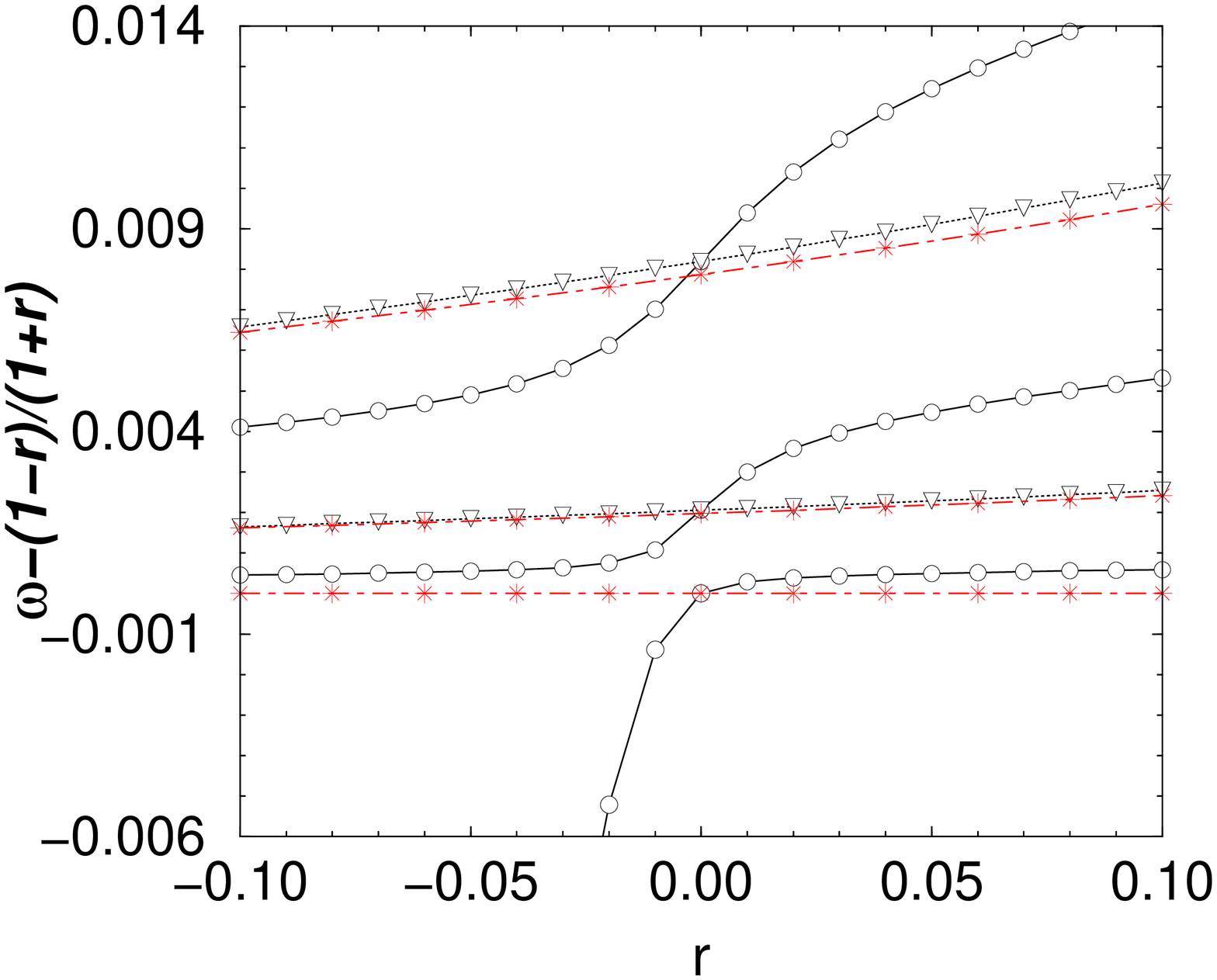,width=1.95in,angle=-90}
\psfig{file=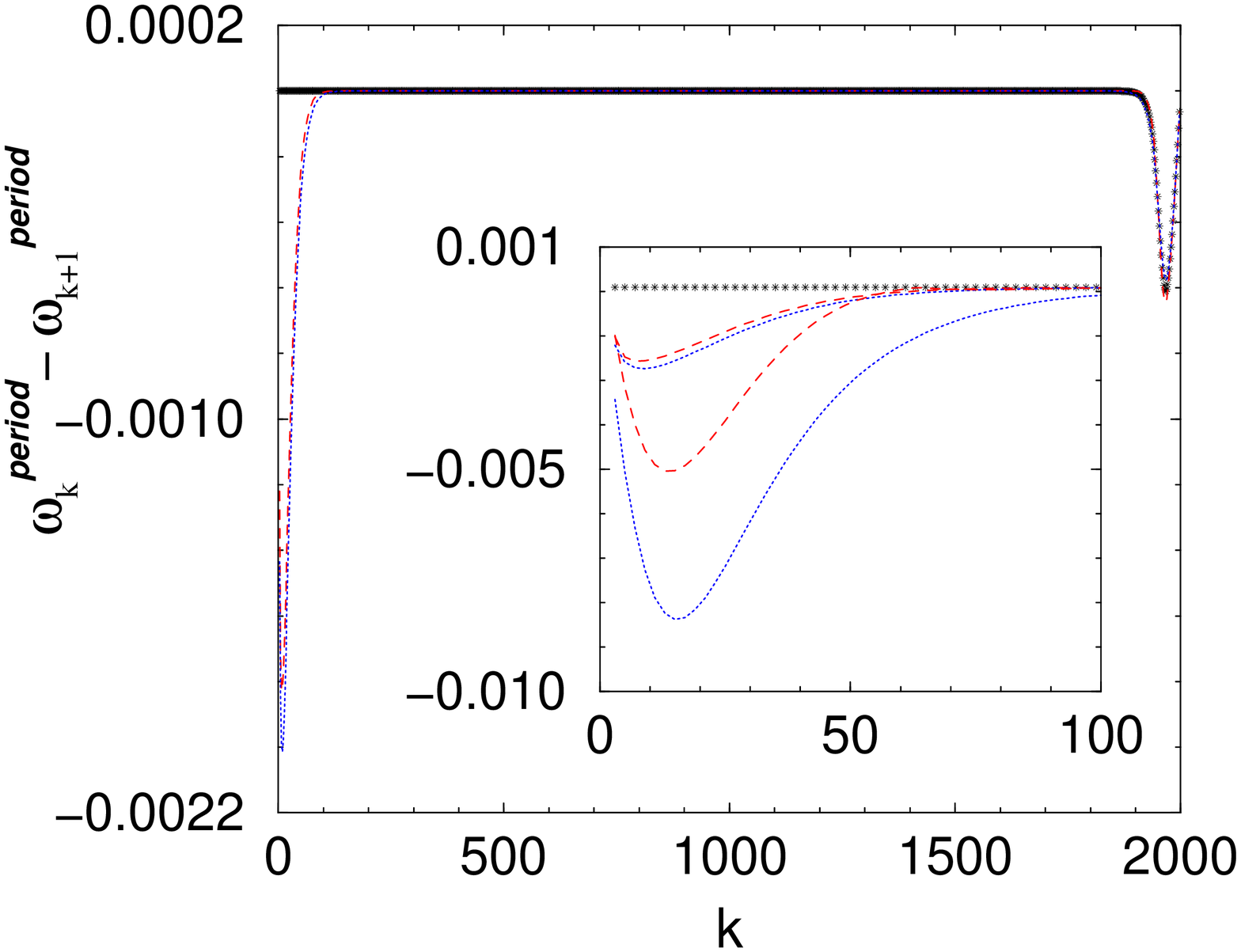,width=1.95in,angle=-90}
\caption{Periodic BC: 
The left panel shows the difference between the numerical eigenfrequencies 
computed from Eq.\ (\ref{eq3}), 
and $\omega_{min}$ versus 
$r$. The frequencies corresponding 
to the spectrum of the linear PR case are also shown by 
stars for free BC 
and dot-dashed line for the fixed BC case. 
In the right panel we show the difference 
between $\omega_{k}^{period}-\omega_{k+1}^{period}$ versus 
$k$ ($k=2, 4, ..., N-2$).  
The stars practically at zero for all $k$ 
represent the integrable system ($r=0$), 
whereas the dashed ($r=-0.02$) and dotted ($r=0.02$) 
lines correspond to  non-integrable cases.
In the inset we  observe that the difference 
between these frequencies increases as $r$  increases 
(the case of 
$r=-0.1$ (dashed line) and $r=0.1$ (dotted line) are also included). 
}
\label{nfig2}
\end{figure}

\begin{figure}[h]
\centerline{\psfig{file=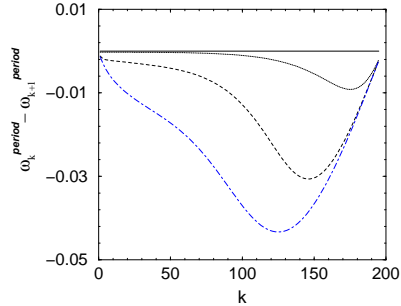,width=1.65in,angle=-90}}
\caption{Periodic BC for AL-DNLS: 
The solid line at zero represents the difference between two 
consecutive frequencies ($k=2, 4, ...$) 
for the integrable A-L lattice ($\epsilon=0$). 
The double multiplicity of the frequencies is destroyed as  
$\epsilon$ is increased (dotted, dashed and dot-dashed lines 
represent the non-integrable cases of $\epsilon=0.1, 0.5, 1$ 
respectively). }
\label{nfig5}
\end{figure}

Our results when the parameter of the PR potential
or $\epsilon$ in the DDE are varied can be summarized in Figs. \ref{nfig1}-\ref{nfig5}.

From the above results the following conclusions can be drawn: \\
1. For fixed BC, the CS band
edge frequency is prohibited. Hence, we compare
$\omega_k^{free}$ with $\omega_{k-1}^{fixed}$. 
We find that {\it for small wavenumbers}, fixed and free
BC eigenfrequencies practically coincide {\it only} 
in the integrable case, 
whereas for the non-integrable case 
we observe an oscillatory behaviour of this function.  \\
2. For periodic BC, the CS comprises of modes coming alternately
from the free and fixed BC. This seems natural as the free
boundary conditions select eigenmodes symmetric at the boundary,
the fixed ones select modes antisymmetric at the boundary,
while the periodic BC allow for both. \\
3. An additional feature, equally important as 1. (especially in
view of its potential predictive power) is the
fact that for the integrable case of $r=0$, periodic BC
essentially imply the presence of {\it double} eigenvalues.
The difference between the two eigenvalues is
 of the $O(10^{-9})$  for all pairs
(except for the cutoff, discretization induced 
phenomena at the upper end of the spectrum).
This is in sharp contrast (in particular for small wavenumbers),
to even mild breakings of integrability, as can be inferred from
Fig. \ref{nfig2}. \\
4. Statements 1. and 3. above can be used in predictive form and
constitute the criterion (algorithm) set forth in this work: for
a given PDE/DDE model, find the steady state coherent structure
(e.g., by finding 
the exact solution of an ODE, or numerically
performing a Newton type algorithm). Linearize around the
exact solution and study, in particular, the small wavenumbers,
close to the lower edge of the CS (we assume that the problem is
monoparametric in what follows, but it is clear that the application
of the criterion does not require that). If for a
critical/singular value of the parameter the fixed BC and free BC
(small $k$)
eigenvalue spectra essentially coincide and the multiplicity of
periodic BC eigenvalues becomes double, 
then the model for this
unique value of the parameter can be ``strongly suspected'' to be 
integrable. We use the above expression, as we provide no rigorous
proof, but only supporting (but rather universal in distinct models
with distinct features/solutions) numerical evidence for this
statement. \\
5. We now attempt to give {\it
a qualitative explanation} for the criterion set forth.
One can solve explicitly (by means of Chebyshev polynomials) 
the discretized linearization problem of the 
PR model, to find:
$\omega_{k+1}^{free} =  \sqrt{\omega_{min}^{2} + 
({4}/{(\Delta x)^{2}}) \sin^{2}({k \pi}/
{ (2 N-2)})},  
\quad k=0, 1, ..., N-2$; 
for fixed BC  
$\omega_{k}^{fixed} = \sqrt{\omega_{min}^{2} + 
({4}/{(\Delta x)^{2}}) \sin^{2} ({k \pi}/
{(2 N)})}, 
\quad k=1, 2, ..., N-1$; 
for periodic BC
$\omega_{0} =  \omega_{min}, \qquad \omega_{k} =  \sqrt{\omega_{min}^{2} + 
({4}/{(\Delta x)^{2}}) \sin^{2} (k \pi /(N-1)) },  
\qquad \omega_{N-1-k} =  \omega_{k}, 
\quad k=1, ..., (N-2)/2,   $
respectively. From the above expressions,
$\omega_{k}^{free} \approx \omega_{k-1}^{fixed}$ for small $k$. 
The 
difference between these two sets of frequencies as a function of $k$ 
practically  
does not depend on $r$ and coincides with the spectrum for the 
nonlinear integrable case $r=0$, except for the oscillatory behavior 
at the edge. 
For periodic BC, except for the first phonon frequency, 
$\omega_{k+1} = \omega_{N-2-k}$, all the eigenfrequencies 
are of algebraic multiplicity $2$. 
In both situations (for fixed-free and for periodic BC), the 
linearization around the nonlinear wave (kink) shows the
{\it exact same features} as the linearization around the
uniform steady state {\it only in the integrable case}. That
is because, it is only in that case, that the potential
is {\it reflectionless} \cite{AS,FAD} and, hence, there
is no reflection in the scattering from the nonlinear wave.
Such a reflection would lead to an eigenfrequency adjustment
in the non-integrable case (in order to satisfy the boundary
value eigenvalue problem). However, it is only for the 
integrable case that
the CS (used here to ``perceive'' the presence of the nonlinear
wave through the scattering) will not ``feel'' the wave
(as only in this singular case there is no reflection from it).
\\
7. We have also tested the validity of
these results also in 
equation (\ref{neq2}), in the vicinity of the integrable limit $\epsilon=0$,
with similar conclusions.
For brevity, we only
show the case of periodic BC.
In Fig. \ref{nfig5}, it can be clearly seen that it is only
for the integrable case that the double eigenvalue
multiplicity is obtained. 


In conclusion, we have proposed and used a new 
test for revealing the potential integrable nature of a given
model problem. By varying the boundary conditions and ``feeling''
the effects of such variations in the CS,
we have revealed that the small wavenumbers (particularly, but
also the CS more generally) have  singular
ways of responding to the unique parameter values for which the
model is integrable. These singular features (such as an approximate
identification of fixed with free BC, small $k$ eigenvalues and the
double multiplicity of eigenvalues for periodic BC) 
can be used to identify and single out
the integrable behavior. We have provided two model examples 
respectively for kinks and pulses and for a PDE and a DDE.
Independently of the detailed structure of the model these 
properties  have been identified as universal. It 
is naturally interesting to explore the potential usefulness of such
a criterion in various more complex settings.
It would
also be of value to rigorously elucidate the mathematical structure 
that underlies the findings presented herein (we have only 
qualitatively attempted to justify this structure here).

\end{document}